\documentclass[a4paper,11pt]{article}
\usepackage{CAR}
\usepackage[T1]{fontenc}
\usepackage{amsfonts}
\usepackage{amsmath}
\usepackage{amssymb}
\usepackage{todo}

\begin{document}
\setcounter{footnote}{0}
\setcounter{figure}{0}


\Abschnitt
{Post-Quantum Secure Cryptographic Algorithms}
{Post-Quantum Secure Cryptographic Algorithms}
{rubrik}

\vspace{3mm}


\Aufsatz
{Post-Quantum Secure Cryptographic Algorithms}
{Overview of Developments 2017/2018}
{Dipl. Math. Xenia Bogomolec, Dr. Jochen Gerhard}
{NameAuthor}
{Dipl. Math. Xenia Bogomolec, X4pi GmbH\\ Dr. Jochen Gerhard, BearingPoint Software Solutions GmbH}
{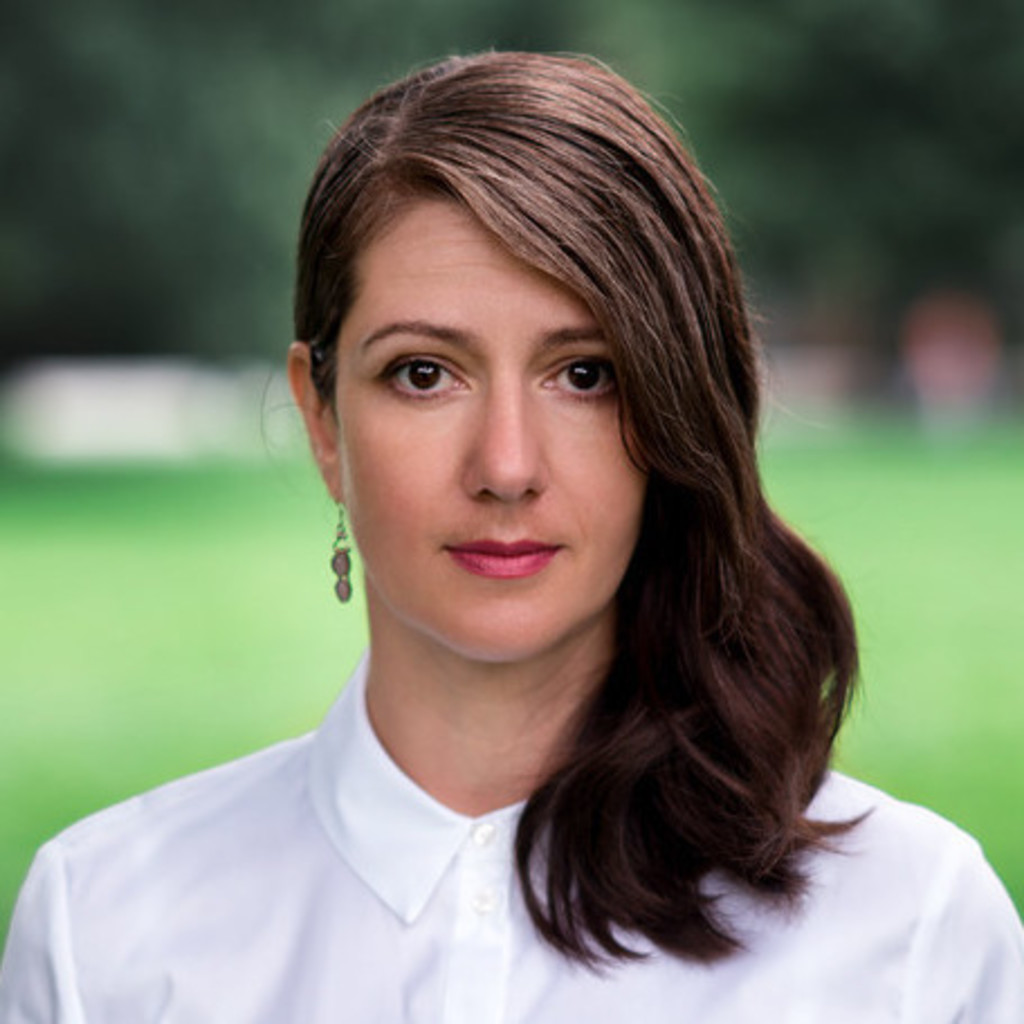,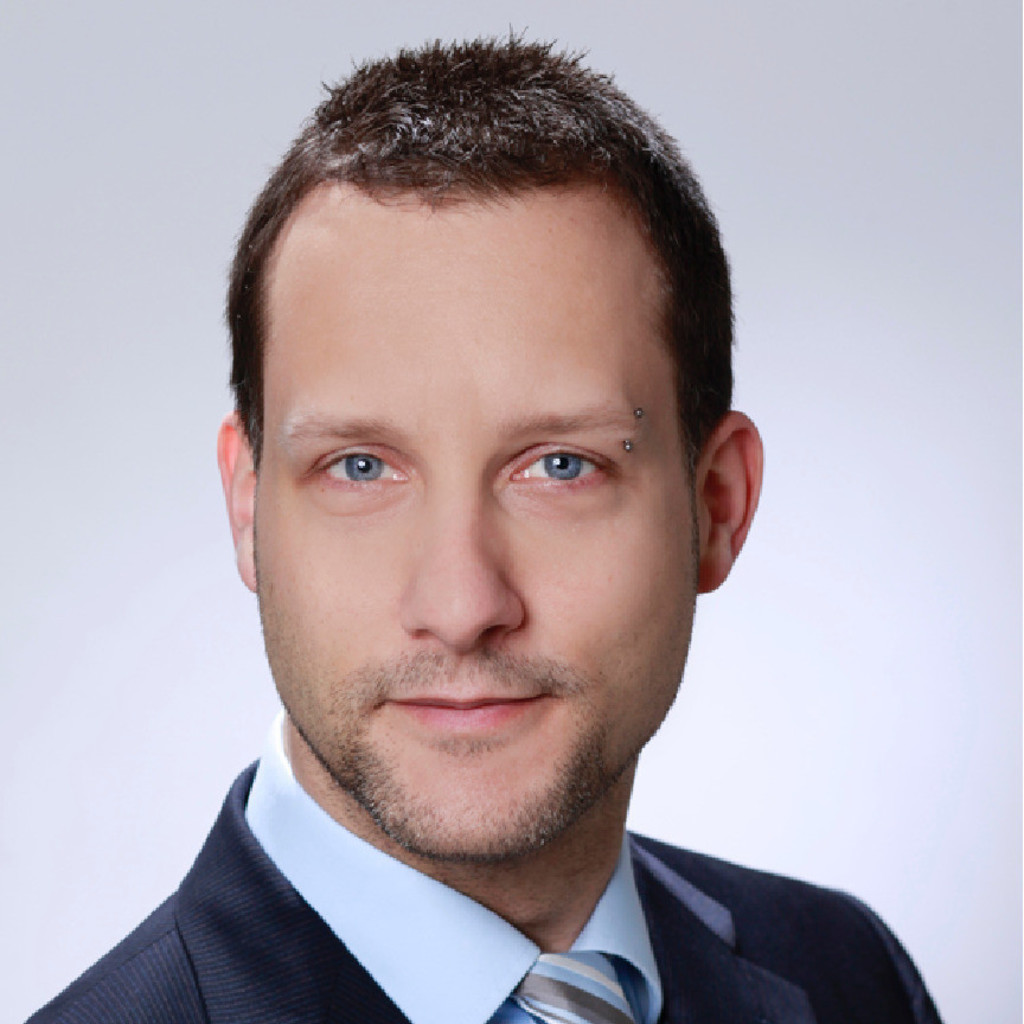}
{indigomind@protonmail.ch,jochen.gerhard@bearingpoint.com}

\begin{otherlanguage}{english}

\vspace{3mm}
\begin{multicols}{2}
\noindent


\Ueberschrift{Introduction}{intro}
The expected dawn of a new technological era has certainly begun when IBM offered their first commercially available 20-Qubit Quantum Computers November 2017.
While it was still discussed if it was necessary to take quantum technology into account in the IT industry during the last year, the estimations about their capability evolvement become much more specific now.\\ \\
Luckily scientific researchers have specialized in the examinations of the various resulting challenges and questions since the beginning of this century. A series of conferences about post-quantum cryptography, the PQCrypto, started in 2006. Since 2010, they take place in another town of the world every year. The following article gives an overview of current developments in algorithmic solutions answering the upcoming threats posed by quantum computers as well as unsolved problems in the classical IT landscape. 

\Ueberschriftu{Quantum Technologies}
Quantum-mechanical phenomena, such as superposition and entanglement, are used for communication, computing, sensoring and simulation. While communication, sensoring and simulation have been realized in publicly announced projects or products, quantum computing was only a matter of research until last november. \\
With the advent of 49 qubit processors quantum supremacy lies within reach, i.e. the potential ability of quantum computing devices to solve problems that classical computers practically cannot solve \cite{IEEESpec, googleai}. IBM has announced to have built a 50 qubit prototype, Google participates in the race with their new record-breaking 72-qubit quantum processor Bristlecone.

\Ueberschriftu{Benefits}
Quantum technologies offer and promise major benefits. So called adiabatic quantum computers, e.g. the D-Wave 2000Q with 2048 qubits from D-Wave Systems in Canada are able to solve optimization problems that would overburden a classical computer. Photon based quantum key distribution devices from ID Quantique in Switzerland are used by the government in Geneva and other institutions. China has built the 2000km quantum communication channel QUESS between Beijing and Shanghai for banks, the Xinhua News Agency and the government, whose nodes receive keys from their quantum communication satellite. Last year they denoted feasible distances up to 1200 km. \\
In the future quantum computers with enough stable qubits are expected to be able to help building complex materials as well as solve medical and environmental problems amongst other things.

\Ueberschriftu{Threats}
It is long known that the security of currently used cryptographic algorithms relying on the hardness of integer factorization and finding discrete logarithms (DLOG systems) \cite{SHO} will expire with potent enough quantum computers. All public parameters like public keys from asymmetric key pairs can then be used to compute the corresponding private keys. With the knowledge of those private keys, encrypted data which was collected and assigned to the relevant key exchanges, will no longer remain secret. For technologies like public distributed ledgers, where encrypted data is publicly available, this threat is even more serious. \\

\Ueberschrift{Solutions}{solutions}

\Ueberschriftu{Quantum Key Distribution}
QKD is an implemented cryptographic protocol for key distribution involving components of quantum mechanics. The security of encryption that uses quantum key distribution relies on the foundations of quantum mechanics. In this context, the process of measuring a quantum system in general disturbs the system itself. So any third party trying to gain knowledge of the key would be detected by the original communication parties. \\ 
\\
Quantum key distribution networks have already been established in China (QUESS), Austria (SECQC), Japan (Tokyo QKD Network), Switzerland (SwissQuantum) and the USA (DARPA). 
Disadvantages for widespread practical usage are limited distances between communication partners and the need of expensive hardware. \\
Rarely mentioned is the fact that message source authentication does not come with QKD genuinely. Man in the middle attacks are also possible if the communication parties do not agree on an authentication protocol beforehand.  

\Ueberschrift{Post-Quantum Cryptography}{pqcrypto}
\noindent
The alternative to QKD are algorithms whose security rely on mathematical properties, like hardness of computing the inversion of a one way function even with a quantum computer. There are four mathematical areas which offer solutions for encryption, key exchanges and signatures. Some of them are still in the middle of the research process, others have been observed and challenged for years. The advantages of post-quantum cryptography are that they can run effectively on currently used devices such as smart phones, desktops and IoTs and they can be enabled by simple software updates.

\Ueberschriftu{Code-Based}
Syndrome decoding of linear error-correcting codes are NP-complete considered as a decision problem if the number of errors are unbounded. On the other hand, some classes of linear codes have very fast decoding algorithms. The basic idea of a code-based crypto system is to choose a linear code with fast decoding algorithm and disguise it as a general linear code. Then the attacker has to use syndrome decoding for decrypting the message while the message receiver, who also set up the system, can remove the disguise and use the fast decoding algorithm. \\
\\
\textsc{McEliece} and the \textsc{Niederreiter} cryptosystems are two basic encryption schemes built on this setup. \textsc{McEliece} was the first scheme using randomization in the encryption process. Both systems consist of three algorithms:

\begin{itemize} [noitemsep, nolistsep]
\item[1)] Probabilistic key generation algorithm producing an asymmetric key pair
\item[2)] Probabilistic encryption algorithm
\item[3)] Deterministic decryption algorithm
\end{itemize} 

\vspace{0.4cm}

The private key is an $(n,k)$-linear error correcting code represented by a generator matrix $G$, with a known efficient decoding algorithm. Originally binary Goppa Codes with the Patterson decoding algorithm were used. The public key is the generator matrix $G$ perturbated by two randomly chosen invertible matrices $S$ and $P$

$$ G^\prime = SGP $$
\\
where $S$, a $(k \times k)$ matrix, functions as a scrambler and $P$ is a $(n \times n)$ permutation matrix. Parameters proposed by \textsc{McEliece} \cite{MCE} result in a public key of $2^{16}$ bytes size. The most effective attacks on \textsc{McEliece} use information-set decoding. To resist those in a quantum computing context, key sizes have to be increased by a factor of 4.\\
\\
The \textsc{Niederreiter} scheme \cite{NR} applies the same idea to a parity check matrix $H$ of a linear code. The encryption is about ten times faster than McEliece. McEliece was originally believed not to be usable for authentication or signature schemes because the encryption algorithm is not one-to-one and the total algorithm is truly asymmetric, meaning, encryption and decryption do not commute. However, a one-time signature scheme based on \textsc{McEliece} and \textsc{Niederreiter} was 
proposed at the Asiacrypt in $2001$ \cite{CFS}: \\

\begin{itemize} [noitemsep, nolistsep]
\item[1)] Choose a hash function $h$  and compute the hash value $h(d)$ of the document $d$ which has to be be signed
\item[2)] Decrypt the hash value $h(d)$ as if it was an instance of the ciphertext
\item[3)] Append the decrypted hash value to the document as a signature
\end{itemize} 

\vspace{0.4cm}
\noindent
As the second step in the signature scheme almost always fails, the system additionally specifies a deterministic way of tweaking $d$ until a hash value $h(d)$ is found which can be decrypted. Verification then applies the public encryption function to the signature to the signature and compares it to the hash value of the document. \\
\\
The most recently published code-based key exchange protocol is \textsc{Ouroboros} \cite{OUR}. It uses quasi-cyclic codes in Hamming metric in the encryption algorithm, efficient decoding is achieved through bit flipping in the Random Oracle Model. Encryption and decryption are faster than RSA for comparative benchmarks (https://bench.cr.yp.to). Ouroboros' integration into the OpenSSL/TLS library is planned and it is proposed as post-quantum secure algorithm at the NIST.

\Ueberschriftu{Hash-Based}
This domain is limited to digital signatures schemes which rely exclusively on the security of the underlying hash functions so far. The signatures themselves reveal a part of the signing key and can only be used for one message, same as it is known from one-time pads such as visual cryptography shares. \\
\\
Merkle tree signature schemes, introduced in 1979, combine a one-time signature scheme with a Merkle tree structure. Building blocks of the Merkle trees are one-time signature key pairs, with the node at the top being the global public key. This typically 256 bit large key can be verified with the path to another given public one-time key in the tree using a sequence of tree nodes, called the authentication path. The global private key is usually derived from a seed generated by a pseudo random number generator and has the size of 256 bits as well.
Hereby, the number of possibilities for such signatures are all possible combinations of the simple one-time signatures within the tree structure. This procedure considerously enhances the security of the scheme against brute force attacks. \\
\\
The latest performance improved hash-based signature scheme is \textsc{SPHINCS\textsuperscript{+}}\cite{SPH+}, the advanced \textsc{SPHINCS} \cite{SPH} scheme which was presented at EUROCRYPT 2015. Unlike its predecessors, XMSS and LMS, it is stateless, meaning that signing doesn't require updating the secret key. It is a so called few-times scheme, where "few-times" means as much as after $2^{64}$ signatures it is necessary to reinitiate the complete scheme. Its signature sizes range from 8kb for NIST security level 1 to 30kb for NIST security level 5.

\Ueberschriftu{Lattice-Based}
Lattice based codes come with the challenge of finding the nearest lattice point or a shortest basis for a given lattice. Both problems and their approximate adequates have been solved with NP-hard algorithms only. Given they are one of the longest known public key crypto systems, they can be fairly seen as the most promising post quantum crypto approaches. Low memory requirements and high speed computations let them run effectively on all currently and widely used devices. However, due to their significantly bigger key sizes they had not been as thoroughly researched and applied as \textsc{RSA}, \textsc{El Gamal} \cite{IMC} or \text{DLOG systems}. \\
\\
\textsc{NTRU} was the first successful lattice-based asymmetric cryptosystem. It was was proposed and patented in 1996 \cite{NTR}. With the expiration of the patent in 2016, \textsc{NTRU} Prime \cite{NTP}, an improvement by eliminating worrisome algebraic structure could be published. Their security rely on the interaction of a polynomial mixing system with the independence of reduction modulo two relatively prime integers $p$ and $q$. \\
\\
Another popular ingredient of lattice-based algorithms is the Learning with Errors (LWE) problem. It was used in \textsc{BCNS} \cite{BCN}, which phrased Peikerts key encapsulation algorithm as a key exchange protocol. \textsc{BCNS} was the first lattice-based algorithm which was integrated into the OpenSSL library. \\
\\
With \textsc{New Hope} \cite{NHP} an improvement was achieved by chosing more efficient parameters and shifting from LWE to Ring Learning with Errors (RLWE). The \textsc{New Hope} protocol allows man in the middle attacks, message authentication has to be implemented additionally. Google ran an experiment by using \textsc{New Hope} embedded in an ECC procedure for a certain number of connections between the Chrome browser and their own servers in 2016. Since 2017, Infineon works on the first generation of contactless post-quantum chips with Pöppelmann, one of the authors of the \textsc{New Hope} paper. \\
\\
\textsc{Dilithium} \cite{DLM}, a module-lattice-based signature scheme was designed with the intention to be easy to implement against side-channel attacks, while offering comparable efficiency to previously developed lattice-based signature schemes. The key innovation is the replacement of Gaussian sampling by uniformly random sampling over a bounded domain. Furthermore, the public key sizes are reduced by more than a factor of 2.\\
\\
All these algorithms except \textsc{BCNS} are submitted to the NIST post-quantum cryptography standardization process.

\Ueberschriftu{Multivariate}
\noindent
The proven NP-hardness and NP-completeness of solving multivariate polynomial equations over a finite field $F$ are the reason why schemes with those asymmetric cryptographic primitives are considered good candidates for post-quantum security. Most of the published schemes use multivariate quadratics, namely polynomials of degree two. \\
The basic scheme consists of two affine transformations 
$$S: F^n \rightarrow F^n$$
$$T: F^m \rightarrow F^m$$
and an easy to invert quadratic map 
$$P^{\prime} : F^m \rightarrow F^n$$ 
\noindent
The trapdoor $(S^{-1}, P^{\prime^{-1}}, T^{-1})$ represents the private key, whithout which the public key $P = S \circ P^{\prime} \circ T$ is assumed to be hard to invert.\\ 
A first multivariate quadratic scheme, $C^*$ \cite{MI}, was presented at the \textsc{Eurocrypt Conference} 1988. After it was broken \cite{P95}, the general principal was used for stronger schemes, such as \textsc{Hidden Field Equations} \cite{HFE} and \textsc{QUAD} \cite{BGP}. \\
\\
Multivariate signature schemes provide the shortest signatures amongst post-quantum algorithms (\textsc{Gui} \cite{GUI} 129 bit over $GF(2)$ for a quantum security level of 80 bit). The signature $x$ of a message $m$ is created by hashing $m$ into a vector $y \in F^n$ and computing $x = P^{-1}(y) = T^{-1} (P^{\prime}(S^{-1}(y)))$. The receiver can simply compute the hash $y$ and check if $P(x) = y$. \\
\\
\textsc{Medium Field Signature Schemes} \cite{HMFE} with fewer equations and variables in the public key offer  a further reduction in key sizes, greater efficiency and scalable levels of security. A proposal is submitted to the NIST standardization process of post-quantum signature schemes.

\Ueberschriftu{Isogeny-Based}
One of the latest and most challenging post-quantum crypto ideas is the application of isogeny based encryption schemes like \textsc{Supersingular Isogeny Diffie-Hellmann} (\textsc{SIDH}). With 2688-bit public keys at a 128-bit quantum security level, this scheme uses the smallest keys amongst post-quantum key exchanges. Additionally it supports perfect forward secrecy, a property which preserves the confidentiality of old communication sessions even if long-term keys have been compromised. \\
\\
Although they are not as thoroughly researched as the previously mentioned schemes, Microsoft published an experimental VPN-library with a \textsc{Supersingular Isogeny Key Encapsulation} algorithm (\textsc{SIKE}) based on \textsc{SIDH} amongst a \textsc{LWE} key exchange and a signature algorithm using symmetric-key primitives and non-interactive zero-knowledge proofs \cite{MSR}. \textsc{SIKE} is also submitted to the NIST standardization process of post-quantum cryptography schemes\\
\\
In a youtube video of a Microsoft research session where \textsc{SIKE} is presented to other researchers by Christophe Petit, he states at the end: "I wouldn't bet national security on it". On the other hand, \textsc{SIDH} was also denoted as "the hottest thing we have" in the key note of the pqcrypto conference 2017. \\

\Ueberschriftu{Amendment}
Paramater choices are much more delicate for post-quantum crypto schemes than they are for classical ones. Furthermore classical asymmetric schemes mostly rely on number theory, a topic which has been studied in early courses at universities, where post-quantum algorithms include more mathematics from courses which are usually taught at later stages of study courses. \\
\\
It will not only be a challenge to distinguish and weigh the complex influences on security of post-quantum encryption schemes, there will also be an increased need of cooperations between mathematicians, computer scientists and programmers to mitigate flaws in implementations, configurations and applications. \\
\\
For someone who is not familiar with the concept of a mathematical conjecture, it is hard to understand on what ground the security of cryptography is built and what time can do to it, with or without regard to emerging technologies. Who can say for sure that there is no-one who generates one RSA key pair after another since decades and stores them in a huge database where he can simply assign a private key to its public key if it is present in his own collection? How many distinctive usable key pairs can even be expected within the range of a 4096-bit integer? 

\vspace{1cm}




\end{multicols}

\end{otherlanguage}

\end{document}